\documentclass{physeauth}
\usepackage{graphicx}
\usepackage{amsmath}
\usepackage{amssymb}

\begin{document} 
\begin{frontmatter}

\title{Parity effect and spontaneous currents in superconducting nanorings   } 
\author[tfp,ipm]{Sergei V. Sharov}
and
\author[int,fian]{Andrei D. Zaikin \thanksref{thank1}}

\address[tfp]{Institut f\"ur Theoretische Festk\"orperphysik,
Universit\"at Karlsruhe, 76128 Karlsruhe, Germany\\   }
\address[ipm]{Institute for Physics of Microstructures, Russian Academy of
Sciences, 603950 Nizhny Novgorod, Russia,\\}
\address[int]{Forschungszentrum Karlsruhe, Institut f\"ur Nanotechnologie,
76021 Karlsruhe, Germany,\\}
\address[fian]{I.E.Tamm Department of Theoretical Physics, P.N.Lebedev
Physics Institute, 119991 Moscow, Russia\\}
\thanks[thank1]{Corresponding author.  E-mail: zaikin@int.fzk.de}

\begin{abstract}
 New physical effects emerge from an interplay between
the electron parity number and persistent currents in
superconducting nanorings. An odd electron, being added to the ring,
produces a countercurrent which may substantially modify the ground state
properties of the system. In superconducting nanorings
with an embedded normal metal layer a novel ``$\pi /N$-junction'' state can
occur for the odd number of electrons. Changing this number from even to
odd yields {\it spontaneous} supercurrent in the ground state of
such rings without any externally applied magnetic flux. Further peculiar
features of the parity effect are expected in structures with resonant
electron transport across the weak link. 
\end{abstract}

\end{frontmatter}

\section{Introduction}
It is well established both theoretically and experimentally
\cite{AN92,Tuo93,Laf93,JSA94,GZ94,AN94} that thermodynamic properties of
isolated superconducting systems are
sensitive to the parity of the total number of electrons $\mathcal{N}$
despite this number may be macroscopically large. The fundamental difference
between superconductors with even and odd $\mathcal{N}$
is due to the fact that the number of electrons in the
condensate of Cooper pairs should necessarily be even.
Hence, for odd $\mathcal{N}$ at least
one electron remains unpaired having an extra energy equal to the
superconducting gap $\Delta$. This difference --
hindered by the entropy factor at sufficiently high temperatures -- becomes
important in the low temperature limit.

Can the supercurrent be affected by this parity effect? Recently
we have argued \cite{SZ} that (a) the answer to this question is
positive and (b) in superconducting nanorings interrupted by a
weak link parity-restricted supercurrent can strongly deviate from
one evaluated for the grand canonical ensemble. In this paper we
will extend our previous analysis focusing it on superconducting
rings with embedded piece of a normal metal ($SNS$ rings).
We will consider transparent nanojunctions as well as structures with low
transmission of $NS$-interfaces. In the latter case resonant effects
become important leading to substantial modifications
of the physical picture of the parity effect \cite{SZ}.

\section{Parity projection formalism}

In the subsequent analysis of parity-restricted persistent currents (PC) in
superconducting nanorings we will employ the parity projection
formalism \cite{JSA94,GZ94,AN94}. Recapitulating the key points of
this approach we will closely follow Ref. \cite{GZ94}. 
We will then derive a general formula
for the circulating supercurrent in isolated superconducting rings with 
even/odd number of electrons.

The grand canonical partition function $\mathcal{Z}(T,\mu )= {\rm Tr
} e^{-\beta(\mathcal{H}-\mu \mathcal{N})}$ is linked to the canonical
one $Z(T,\mathcal{N})$ as
\begin{equation}
\label{6}\mathcal{Z}(T,\mu )=\sum\limits_{\mathcal{N}=0}^\infty Z(T,\mathcal{N})\exp \biggl({\frac{%
\mu \mathcal{N}}T}\biggr).
\end{equation}
Here and below $\mathcal{H}$ is the system Hamiltonian
and $\beta \equiv 1/T$. Inverting
this relation and defining the canonical partition functions
$Z_{e}$ and $Z_{o}$ respectively for even ($\mathcal{N}\equiv 
\mathcal{N}_{e}$) and odd ($\mathcal{N} \equiv \mathcal{N}_{o}$) ensembles, one
gets
\begin{equation}
Z_{e/o}(T)={\frac 1{2\pi }}\int\limits_{-\pi }^\pi due^{-i\mathcal{N}_{e/o}u}%
\mathcal{Z}_{e/o}(T,iTu),
\label{invrel}
\end{equation}
where
$$
\mathcal{Z}_{e/o}(T,\mu )={1\over 2} {\rm Tr }\left\{\big[1\pm (-1)^{\mathcal{N}}\big]
e^{-\beta(\mathcal{H}-\mu \mathcal{N})}\right\}
$$
\begin{equation}
\label{11}={\frac 12}\bigl(\mathcal{Z}(T,\mu )\pm \mathcal{Z}(T,\mu +i\pi T)\bigr)%
\end{equation}
are the parity projected grand canonical partition functions. For
$\mathcal{N} \gg 1$ it is sufficient to evaluate the integral in
(\ref{invrel}) within the saddle point approximation
\begin{equation}
\label{14}Z_{e/o}(T)\sim e^{-\beta (\Omega _{e/o}-\mu _{e/o}
\mathcal{N}_{e/o})},
\end{equation}
where 
$${\Omega }_{e/o} =-T\ln \mathcal{Z}_{e/o}(T,\mu )
$$ 
are the
parity projected thermodynamic potentials,
\begin{eqnarray}
{\Omega }_{e/o}={\Omega }_{f} - T \ln \bigg[ \frac{1}{2} \Big( 1
\pm e^{- \beta ({\Omega_{b}} - {\Omega_{f}})} \Big) \bigg]
\label{Ome/o}
\end{eqnarray}
and 
$$
{\Omega}_{f/b}= - T \ln \left[ {\rm Tr }\left\{(\pm1)^{\mathcal{N}}
e^{-\beta(\mathcal{H}-\mu \mathcal{N})}\right\} \right].
$$
``Chemical potentials'' $\mu _{e/o}$ are
defined by the saddle point condition
$$
\mathcal{N}_{e/o}=-\partial \Omega _{e/o}(T,\mu_{e/o})/\partial \mu_{e/o}.
$$

Let us emphasize that $\Omega_{f}$ is just the standard grand
canonical thermodynamic potential and $\Omega_{b}$ represents the
corresponding potential linked to the auxiliary partition function 
$\mathcal{Z}(T,\mu +i\pi T)$. This function can be conveniently
evaluated by finding the
true grand canonical partition function $\mathcal{Z}(T,\mu )$,
expressing the result as a sum over the Fermi Matsubara
frequencies $\omega_f =2\pi T(l+1/2)$ and then substituting the
Bose Matsubara frequencies  $\omega_b =2\pi Tl$ ($l=0,\pm 1,...$)
instead of $\omega_f$.

Consider now an isolated superconducting ring pierced by the
magnetic flux $\Phi$. Making use of the above expressions one
one can easily determine the
equilibrium current $I_{e/o}$ circulating inside the ring:
\begin{equation}
I_{e/o}=I_{f}\pm \frac{I_{b}-I_{f}}{ e^{\beta ({\Omega_{b}} -
{\Omega_{f}})} \pm 1}, \label{Ie/o}
\end{equation}
where the upper/lower sign corresponds to the even/odd ensemble
and we have defined
$$
I_{e/o}= -c \left( \frac{\partial \Omega_{e/o}}{\partial \Phi}
\right) _{\mu(\Phi)},\;\;\; I_{f/b}= -c \left( \frac{\partial
\Omega_{f/b}}{\partial \Phi} \right) _{\mu(\Phi)}.
$$
Below we will make use of these equations in order to
evaluate PC in superconducting nanorings.

\section{Superconducting rings with nanojunctions}

\subsection{The model and general analysis}

Before we turn to concrete calculations let us specify the model
for our system. In what follows we will investigate PC in
superconducting nanorings with cross section $s$ and perimeter
$L=2\pi R$. Superconducting properties of such rings will be
described within the (parity projected) mean field BCS theory. A
necessary -- though not always sufficient -- validity condition
for this mean field approach reads $N_{\rm r} \sim p_F^2s \gg 1$,
i.e. the total number of conducting channels in the ring $N_{\rm
r}$ should remain large. More detailed requirements can be formulated with 
the aid of the results \cite{QPSth,QPSth2}. For generic wires QPS effects 
are small provided the parameter  $\sqrt{s}$ exceeds $\sim 10$ nm
\cite{QPSth,QPSth2,QPSexp,QPSexp2}. Considering then homogeneous rings one
can prove \cite{SZ} that in such rings the effect of the electron
parity number on PC remains small in the parameter $\sim 1/N_{\rm
r} \ll 1$. On the other hand, in rings with only few number of
channels \cite{Kang,Yak} $N_{\rm r}\sim 1$ fluctuation effects
may entirely wipe out the supercurrent thus making the mean field
approach obsolete.

The way out is to consider superconducting rings with $N_{\rm
r}\gg 1$ interrupted by a weak link with only few number of
conducting channels $N$. In such systems the mean field BCS
description remains applicable and, on the other hand, the parity
effect can be large due to the condition $N \sim 1$. Quantum point contacts
(QPC) or, more generally, $SNS$ junctions (of length $d$) can be used for
practical realization of such weak links. The corresponding structure is
depicted in Fig. 1.

\begin{figure}[tb] 
\begin{center}
   \includegraphics[angle=0,width=.5\textwidth]{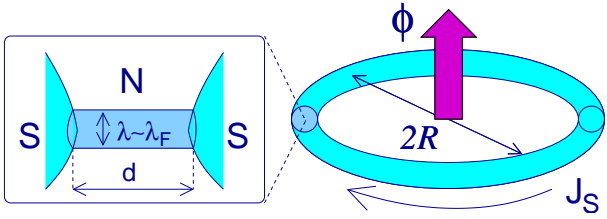}
\caption{Superconducting ring with embedded $SNS$ junction of length $d$.} 
  \end{center} 
\end{figure}

In this case the thermodynamic potential of
the system $\Omega$ consists of two different contributions \cite{FN9}
\begin{equation}
\Omega =\Omega^{(r)}(\mu , T, \Phi_x,
\varphi)+\Omega^{(c)} (\mu , T, \varphi) \label{qpc}
\end{equation}
respectively from the bulk part of the ring and from the contact. The
optimal value of the phase difference $\varphi$ across the weak link is
fixed by the condition $ \partial \Omega /\partial \varphi
=0$ which reads
\begin{equation}
-c\frac{\partial \Omega^{(r)}}{\partial
\Phi_x}=-\frac{2e}{\hbar} \frac{\partial
\Omega^{(c)}}{\partial \varphi}. \label{eqv}
\end{equation}
In (\ref{eqv}) we made use of the fact that the thermodynamic potentials of
the ring depend both on $\Phi_x$ and $\varphi$ only via the
superfluid velocity $ v_{S}=(1/4\pi mR)(\varphi-2\pi
\Phi_x/\Phi_{0})$, in which case one can put $
\partial/\partial \Phi_x =-(2e/\hbar c)(\partial /\partial \varphi )
$. Here and below $\Phi_{0}$ is the superconducting flux quantum.
The left-hand side of Eq. (\ref{eqv}) represents the current
flowing inside the superconducting ring $I^{(r)}=-c\partial
\Omega^{(r)}\partial \Phi_x \simeq (ev_{F}N_{\rm
r}/L)(\varphi-2\pi \Phi_x /\Phi_{0})$. This value should be equal to
the current across the weak link which is given by the right-hand side of
Eq. (\ref{eqv}). Estimating the maximum value of the latter for a
single channel contact as $2e\mathcal{T}\Delta /\hbar$, we obtain
\begin{eqnarray}
\label{L<}\varphi&\simeq &2\pi \frac{\Phi_x}{\Phi_{0}},\;\;\;\mbox{  if  } \;\; L\ll L^{*},\\
\varphi&\simeq &2\pi n, \;\;\;\;\;\; \mbox{     if  }\;\; L\gg
L^{*}, \label{L>}
\end{eqnarray}
where $L^{*}=\xi_{0}N_{\rm r}/\mathcal{T} \gg \xi_0$. In a more
general case of weak links with $N$ conducting channels in the expression
for $L^{*}$ one should set 
$$
\mathcal{T} \to \sum_n^N\mathcal{T}_n.
$$

In what follows we will consider the most interesting limit $N \ll
N_{\rm r}$ and $L \ll L^{*}$. Due to Eq. (\ref{L<}) in this case
the dependence $I_{e/o}(\Phi_x)$ is fully determined by the
current-phase relation for QPS which can be found by means of Eq.
(\ref{Ie/o}) with $I_{f/b}=-(2e/\hbar)\partial
\Omega^{(c)}_{f/b}/\partial \varphi$. 
We also note that quantum fluctuations of $\varphi$ can
easily be suppressed, e.g. by choosing the ring inductance 
sufficiently small. In accordance with the charge-phase
uncertainty relation the number of Cooper pairs passing through
the junction should strongly fluctuate thereby suppressing
charging effects in the contact area.

Let us now directly evaluate the supercurrent in both even and odd
ensembles. As it was demonstrated above, it is sufficient to find
the currents $I_{f/b}$ along with the difference of thermodynamic
potentials $\Omega_b-\Omega_f\equiv \Omega_{bf}$ and combine the
results with Eq. (\ref{Ie/o}). The value $I_{f/b}$ can be
determined by means of a general formula \cite{GZ}
\begin{equation}
I_{f/b}=\frac{2e}{\hbar}\sum_{i=1}^N T\sum_{\omega_{f/b}}
\frac{\sin\varphi}{\cos\varphi + W_i(\omega_{f/b})}, \label{fbfb}
\end{equation}
and $\Omega_{bf}$ is evaluated as a sum of the ring
and the weak link contributions,
$\Omega_{bf}=\Omega_{bf}^{(r)}+\Omega_{bf}^{(wl)}$. 
The term $\Omega_{bf}^{(wl)}$ is found by integrating $I_{f/b}(\varphi
)$ over the phase $\varphi$,
while the ring contribution $\Omega_{bf}^{(r)}$ at low enough $T$
is defined by the standard expression
\cite{Tuo93,JSA94,GZ94} $\Omega_{bf}^{(r)}\simeq \nu \mathcal{V}
\Delta^{1/2}T^{3/2}\exp (-\Delta /T)$, where $\nu$ is the
density of states at the Fermi level, $\mathcal{V}=sL$ is the ring volume
and $\Delta$ is the superconducting energy gap.

\subsection{Rings with quantum point contacts}

Let us first consider the limit of short QPC with $d \to 0$. In this case
one has $W_i(\omega )= (2/\mathcal{T}_i)(1+\omega^2/\Delta^2)-1$,
where $\mathcal{T}_i$ is the transmission of the $i$-th channel.
Combining this formula with Eq. (\ref{fbfb}), after a simple
calculation from Eq. (\ref{Ie/o}) one finds \cite{SZ}
$$
I_{e/o}=-\frac{2e}{\hbar}\sum_{i=1}^N \frac{\partial
\varepsilon_{i}(\varphi)} {\partial \varphi}
\tanh\frac{\varepsilon_{i}(\varphi)}{2T}
$$
\begin{equation}
\times \left[1\pm \frac{(\coth\frac
{\varepsilon_{i}(\varphi)}{2T})^{2}-1}{e^{\beta\Omega_{bf}^{(r)}}\prod\limits_
{j=1}^N
(\coth\frac{\varepsilon_{j}(\varphi)}{2T}) ^{2} \pm 1}\right].
\label{result}
\end{equation}
where $\varepsilon_{i}(\varphi)=\Delta \sqrt{1-\mathcal{T}_i\sin^2(\varphi/2)}$. The term in the square brackets
is specific for canonical ensembles and should be substituted by one
for grand canonical ones. For $N=1$ and at $T=0$ this term reduces to unity for
even ensembles and to zero for odd ones. In other words, in the latter case
PC will be completely blocked by the odd electron and no current will flow
in the system despite the presence of a non-zero external flux threading the
ring. In QPC with several conducting channels the current through the
most transparent one will be blocked by the odd electron and, hence, the 
effect can remain significant also for QPC with $N>1$.

The physics of this blocking effect was discussed in details in Ref. 7. It is
related to a complete cancellation of the contributions to the supercurrent
coming from a pair of discrete Andreev levels with energies
$E_{\pm} (\varphi)=\pm \varepsilon_{i}(\varphi)$. Indeed, for the even ensemble
at $T=0$ only the lower Andreev level $E_-(\varphi )$ is occupied, while
the upper one $E_-(\varphi )$ remains empty. Accordingly, only the lower
Andreev level contributes to the supercurrent in this case, and the result
coincides with the standard grand canonical expression 
\cite{KO,HKR,Furusaki,Bee}. 
In contrast, for the odd ensemble both levels $E_{\pm} (\varphi)$ are
occupied at $T=0$ and contribute to the supercurrent in equal measure.
These contributions, however, enter with opposite signs and completely
cancel each other thus reducing the net supercurrent down to zero.
We also note that a similar result was recently obtained in Ref. \cite{HE}

It is important to emphasize that the above physical picture is based
on the assumption that the supercurrent is determined {\it only}
by the contributions of two discrete Andreev levels
$E_{\pm} (\varphi)$. This is the case provided (a) 
quasiparticle states with energies above $\Delta$
do not contribute to the supercurrent and (b) there are no more
discrete Andreev levels inside the junction. Both these conditions are
met only for symmetric and extremely short QPC with $d \to 0$. In a
general case $d \neq 0$ at least one of these conditions is violated,
and no exact compensation of the supercurrent by the odd
electron countercurrent is anymore possible. The physics
then becomes even more interesting, as it will be demonstrated below.

\subsection{Rings with transparent $SNS$ junctions}

From now on let us lift the condition $d \to 0$, i.e. consider $SNS$ junctions
with a non-zero thickness of the normal metal. Since the parity effect
becomes more pronounced with increasing transmission of the system, let us
first restrict our analysis to fully transparent $SNS$ junctions.
In this case the function $W_i (\omega )\equiv W(\omega )$ is the same for
all channels. It reads
$$
W (\omega )= \left(\frac{2\omega^2}{\Delta^2}+1\right)
\cosh \left(\frac{2\omega d}{\hbar v_F}\right)
$$
\begin{equation}
+
\frac{2\omega}{\Delta}\sqrt{1+\frac{\omega^2}{\Delta^2}}
\sinh \left(\frac{2\omega d}{\hbar v_F}\right).
\end{equation}
Substituting this function
into (\ref{fbfb}) and repeating the the whole calculation as
above, in the limit $T \to 0$ we obtain
\begin{eqnarray}
I_e=\frac{e\Delta N}{\hbar}\left(\sin\frac{\varphi}{2}-
\frac{2y\sin \varphi}{\pi}\ln \frac{1}{y}\right), \\
\label{SCSe}
I_o=I_e-\frac{e\Delta}{\hbar}\left(\sin \frac{\varphi}{2}
+y\mbox{sgn} \varphi \cos \varphi \right)
\label{SCSo}
\end{eqnarray}
for short $SNS$ junctions $y\equiv d\Delta /\hbar v_F \ll 1$ and
\begin{equation}
I_e=\frac{ev_FN}{\pi d}\varphi ,\;\;\;\;I_o=\frac{ev_FN}{\pi
d}\left(\varphi -\frac{\pi \mbox{sgn} \varphi}{N}\right)
\label{SNS}
\end{equation}
for long ones $d \gg \xi_0 \sim \hbar v_F/\Delta$. 
These results apply for $-\pi <\varphi<\pi$ and should be
$2\pi$-periodically continued otherwise. The term containing 
$\ln (1/y)$ in Eq. (\ref{SCSe}) for $I_e$ is written with the logarithmic
accuracy and is valid for $\varphi$ not too close to $\varphi =\pm\pi$.

At $T=0$ the current $I_e$ again coincides with that
for the grand canonical ensembles \cite{Kulik,Ishii}. At the same time
for odd ensembles we observe no blocking effect anymore, but rather
a non-zero current $I_o$ for any non-zero value of $d$, cf. 
Eqs. (\ref{SCSo}-\ref{SNS}). As before, at $T=0$ the countercurrent produced by
the odd electron exactly compensates the current of the next lower Andreev
level. However, the contribution of continuous spectrum
(in the case $d \ll \xi_0$) or of continuous spectrum {\it and} all other
discrete Andreev levels  (in the case $d \gg \xi_0$) remains uncompensated and
is responsible for the non-zero current $I_o$ (\ref{SCSo}-\ref{SNS}).

It is important to emphasize that such a current causes a jump on the
current-phase dependence at $\varphi=0$, and the direction of
$I_o$ at sufficiently small $\varphi$ is always opposite to that of $I_e$.
This property implies that in the case of odd ensembles the minimum energy
(zero current) state occurs not at $\varphi =0$, but at some other value
of the phase difference. For instance, for $d \gg \xi_0$ the ``saw tooth''
current-phase relation is shifted by the value $\pi /N$ and, hence, 
the minimum Josephson energy (zero current) state is reached at 
$\varphi =\pm \pi /N$. One concludes that the ``$\pi/N$-junction'' state 
is realized in this case. This non-trivial state has a number
of specific features. For instance, in the
particular case $N=2$ the current-phase relation $I_o(\varphi )$
turns $\pi$-periodic. In addition, for any $N>1$ the ground state of 
the system $\varphi =\pm \pi /N$ is a twofold degenerate one
within the interval $-\pi <\varphi <\pi$. Here we note that a 
similar behavior is expected for $SNS$ junctions formed by $d$-wave
superconductors \cite{BGZ}. 

Let us also recall that the $\pi$-junction state can be realized in
$SNS$ structures by driving the electron distribution function in
the contact area out of equilibrium \cite{Volkov,WSZ,Yip}. Here,
in contrast, the situation of a $\pi$- or $\pi /N$-junction is
achieved in thermodynamic equilibrium. Despite this drastic
difference, there also exists a certain physical similarity
between the effects discussed here and in Refs.
\cite{Volkov,WSZ,Yip}: In both cases the electron
distribution function in the weak link deviates substantially from
the Fermi function. It is this deviation which is responsible for
the appearance of the $\pi$-junction state in both physical
situations.

The current-phase relations $I_e(\varphi )$ and $I_o(\varphi )$ 
for arbitrary values of the parameter $y$ can be computed numerically. 
Examples are presented in Fig. 2. In the case of odd ensembles one clearly 
observes the current jump at $\varphi =0$. As it is obvious from 
Eq. (\ref{SCSo}), this feature disappears only in the QPC limit $y \to 0$.

\begin{figure}[tb] 
\begin{center}
   \includegraphics[angle=0,width=.5\textwidth]{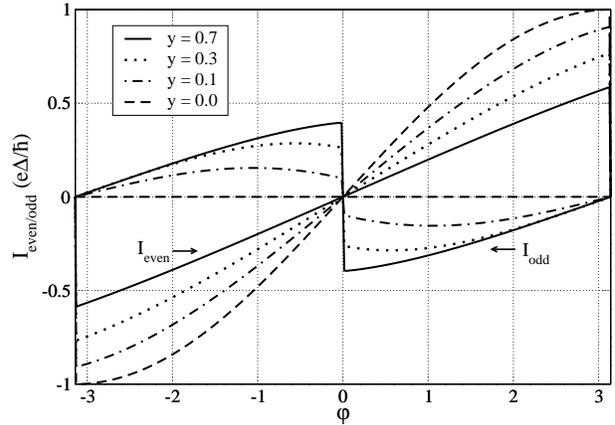}
\caption{The zero temperature current-phase relations $I_e(\varphi )$ and 
$I_o(\varphi)$ ($-\pi <\varphi<\pi$)
 for $N=1$ and different values of the parameter $y=d\Delta /\hbar v_F$.    } 
  \end{center} 
\end{figure}

\subsection{Spontaneous currents in the ground state}

Perhaps the most interesting property of superconducting rings interrupted by a
$\pi$-junction is the possibility to develop {\it spontaneous} 
supercurrent in the ground state
\cite{Leva}. Although this feature is inherent to any type of $\pi$-junctions,
in the case of the standard sinusoidal current-phase relation such
spontaneous supercurrents can occur only for sufficiently large values
of the ring inductance $\mathcal{L}$ \cite{Leva}. In contrast, in the
situation studied here
the spontaneous
current state is realized for {\it any} inductance of the ring
because of the non-sinusoidal dependence $I_o(\varphi )$
(\ref{SNS}) \cite{FN8}.

In order to demonstrate that let us assume that no external flux
is applied to our system. Then at $T \to 0$ the energy of an $SNS$ ring with
odd number of electrons can be written in the form
\begin{equation}
 E_o=\frac{\Phi^2}{2c\mathcal{L}}+
\frac{\pi\hbar v_FN}{\Phi_0^2d}
\left(\Phi -\frac{\Phi_0\mbox{sgn}\Phi}{2N}\right)^2,
\label{ESNS}
\end{equation}
where $\Phi$ is the flux related to the circular current flowing in the ring.
Minimizing this energy with respect to $\Phi$, one easily observes that
a non-zero spontaneous current
\begin{equation}
I=\pm \frac{e v_F}{d}\;\left[1+
\frac{2ev_FN}{d}\frac{\mathcal{L}}{\Phi_0}\right]^{-1}
\label{spcur}
\end{equation}
should flow in the ground state of our system. This is yet one more
remarkable consequence of the parity effect: Just by changing
$\mathcal{N}$ from even to odd one can induce non-zero PC without any
external flux $\Phi_x$. In the limit of small inductances
$\mathcal{L} \ll \Phi_0d/ev_FN$ -- which is easy to reach in the
systems under consideration -- the value of $I$ does not depend
on the number of channels $N$ and is given by the universal
expressions can easily be derived from Eqs. 
(\ref{SCSo}-\ref{SNS}):
\begin{eqnarray}
I_{sp}= e\Delta^2d/\hbar^2 v_F,\;\;\;\; &&\text{if}\;\;d \ll
\xi_0,
\label{sp1}\\
I_{sp}= ev_F/\pi d,\;\;\;\; &&\text{if}\;\;d \gg \xi_0.\label{sp2}
\end{eqnarray}
For intermediate values of the parameter $y$ the amplitude of the
current $I_{sp}$ can be evaluated numerically. The results are
displayed in Fig. 3. In agreement with Eq.
(\ref{sp1}) the spontaneous current $I_{sp}$ increases linearly 
with $d$ at small $d$,
reaches its maximum value $I_{\rm max} \sim 0.4e\Delta/\hbar $ at
$d \sim \xi_0$ and then decreases with further increase of $d$
approaching the dependence (\ref{sp2}) in the limit of large $d$.
For typical BCS superconductors this maximum
current can be estimated as $I_{\rm max} \sim 10 \div 100$ nA.
Spontaneous currents of such a large magnitude can be directly detected 
in modern experiments.

\begin{figure}[tb] 
\begin{center}
   \includegraphics[angle=0,width=.5\textwidth]{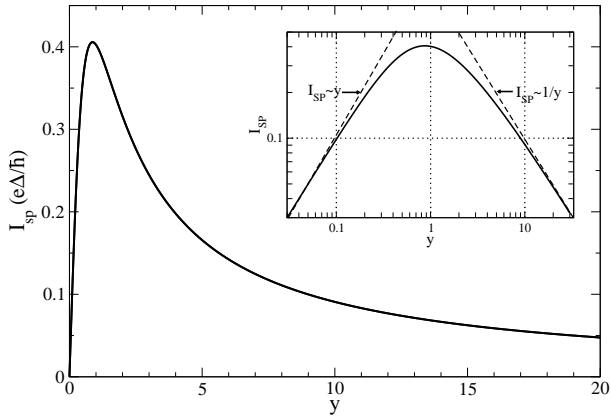}
\caption{The  spontaneous current amplitude $I_{sp}$ as a function of
the parameter $y= d\Delta /\hbar v_F$ at $T=0$. In the inset, the same function is shown on
the $log-log$ scale. Dashed lines indicate the
asymptotic behavior of $I_{sp}(y)$ in the limits of small and large $y$.} 
  \end{center} 
\end{figure}

\subsection{Junctions with resonant transmission}

As we have already emphasized above, the effects discussed here are
mostly pronounced for $SNS$ junctions with few conducting channels $N \sim 1$.
The most promising candidates for practical realization of such structures
are junctions with $N$-layers formed by carbon nanotubes or organic molecules.
In both these cases it is not always easy to achieve good quality contacts
between nanotubes/molecules and superconducting electrodes. In other words,
although the electron transport in the $N$-metal itself can be ballistic,
electron scattering at $NS$-interfaces may be substantial and may
significantly modify the above results.

At the first glance, in the limit of small transmission of $NS$-interfaces
the influence of the parity effect on PC should become weak and, hence, can
be disregarded. Below we will demonstrate that this conclusion -- although
correct under certain conditions -- does not necessarily apply to
nanojunctions for which size quantization of energy levels is an important
effect. This effect gives rize to resonant transmission of electrons across 
the junction and considerably modifies the physical situation.

To proceed we will adopt the standard model of an $SINIS$ junction (``$I$''
stands for the insulating layer) and assume that transmissions of both
$NS$ interfaces are much smaller than unity. In this limit electron transport
across the junction is mainly due to resonant tunneling 
through discrete energy levels inside the normal metal. 
For simplicity we will restrict
our analysis to a single channel junction $N=1$. If needed, generalization
of our analysis to the case $N>1$ can be performed in a straightforward
manner. 

As before, we will make use of the general expression  (\ref{fbfb})
where the function $W(\omega)$ now takes the form \cite{GZ}
\begin{eqnarray}
W (\omega )=\frac{4\sqrt{\mathcal{R}_1\mathcal{R}_2}}{\mathcal{T}_1
\mathcal{T}_2}\frac{\Omega^2}{\Delta^2}
\cos(2k_F d+\phi)
\nonumber
\\+\frac{\Omega^2
(1+\mathcal{R}_1)(1+\mathcal{R}_2)+ \omega^2\mathcal{T}_1\mathcal{T}_2}{\mathcal{T}_1\mathcal{T}_2\Delta^2}
\cosh \left(\frac{2\omega d}{\hbar v_F}\right)\label{W}\\
+\frac{2(1-\mathcal{R}_1\mathcal{R}_2)}{\mathcal{T}_1\mathcal{T}_2}
\frac{\Omega\omega}{\Delta^2}\sinh \left(\frac{2\omega d}{\hbar v_F}\right).
\nonumber
\end{eqnarray}
Here $\mathcal{T}_{1,2}$ and
$\mathcal{R}_{1,2}=1-\mathcal{T}_{1,2}$
are respectively transmission and reflection coefficients of the two $NS$ interfaces, $\Omega=\sqrt{\Delta^2+\omega^2}$ and $ 2k_F d+\phi $ is 
the phase accumulated by electrons during the motion between the barriers.

In accordance with our assumption below we shall consider the limit 
$\mathcal{T}_{1,2} \ll 1$. The most interesting physical situation is
realized in the limit of short $SINIS$ junction $d \ll \xi_0 \sim
\hbar v_F/\Delta$. Under the above conditions the size energy level
quantization in the normal metal becomes an important effect. 
In the case of a one dimensional  metal 
of length $d$, the level spacing in the vicinity of the Fermi energy is 
$\delta \epsilon \sim \hbar v_F/d$. Hence, the condition for the
short junction regime can also be represented in the form $\delta \epsilon
\gg \Delta$. Tunneling of electrons into the reservoirs causes a non-zero
linewidth of the energy levels which is proportional to 
$\mathcal{T}_{1,2}\delta 
\epsilon $. This value is much smaller than $ \delta \epsilon $, 
hence, the resonances remain well separated. In this situation
it suffices to take into account only the
closest to the Fermi energy level inside the normal metal. 

Let us introduce the energy of the resonant level $\epsilon_R$ 
by means of  the relation $\cos(2k_F+\phi)\approx -1+1/2
\left(\frac{ \epsilon_R}{\delta \epsilon }\right)^2$, where $\delta \epsilon
= \hbar v_F/2d$, and the tunneling rates 
$\Gamma _{1,2}/\hbar =\mathcal{T}_{1,2}\delta \epsilon /\hbar$. 
Expanding the $\cosh $ and $\sinh $ terms in Eq. (\ref{W}) to 
the leading order in $\Delta /\delta \epsilon  $ and substituting the result
in Eq. (\ref{fbfb}), we obtain
\begin{equation}
I_{f/b}=\frac{e}{\hbar} T\sum_{\omega_{f/b}}
\frac{\Delta ^2\mathcal{T}\sin\varphi}{\varepsilon ^2(\varphi) + \omega_{f/b}^2
\left(1+4\mathcal{D}\mathcal{T}/\mathcal{T}_{\rm max}   \right) },
\label{fbfb1}
\end{equation}
 where 
\begin{equation}
\mathcal{D}=
\left(\frac{\Delta}{\Gamma}\right)^2\left(1+\frac{
\omega_{f/b} ^2}
{\Delta ^2}\right)+ \frac{\Delta }{\Gamma } \sqrt{1+\frac{\omega_{f/b}^2}
{\Delta ^2}},
\label{fbfb2}
\end{equation}
$\varepsilon (\varphi)=\Delta \sqrt{1-\mathcal{T}\sin^2(\varphi/2)}$,
$\Gamma =\Gamma_1+\Gamma_2 $, $\mathcal{T}_{\rm max}=4 \Gamma_{1} \Gamma_2/
\Gamma ^2$ and the total transmission probability at the Fermi energy
$\mathcal{T}$ is given by the Breit-Wigner formula  
$$
\mathcal{T}=\frac{\Gamma_{1} \Gamma_2}{\left(\epsilon_R \right)^2+\frac{1}{4}
\Gamma ^2}
$$
It is interesting to point out that the expressions defined by 
Eqs. (\ref{fbfb1},\ref{fbfb2}) can also be 
derived from the  Anderson model for resonant tunneling through a 
single impurity center
between two superconductors in the limit of zero on-site interaction.

It follows from Eqs. (\ref{fbfb1}, \ref{fbfb2}) that -- although the
transparencies 
of both barriers are low -- the total transmission $\mathcal{T}$ and, 
hence, the
Josephson current show sharp peaks provided the Fermi energy becomes close to 
the  bound states inside the junction. On the other hand, Eqs. (\ref{fbfb1},
\ref{fbfb2}) demonstrate that even in the vicinity of resonances the 
behavior of the
Josephson current as a function of the phase difference $\varphi$ and
temperature $T$ can substantially deviate from that for
transparent $SNS$ junctions.

\begin{figure}[tb] 
\begin{center}
   \includegraphics[angle=0,width=.5\textwidth]{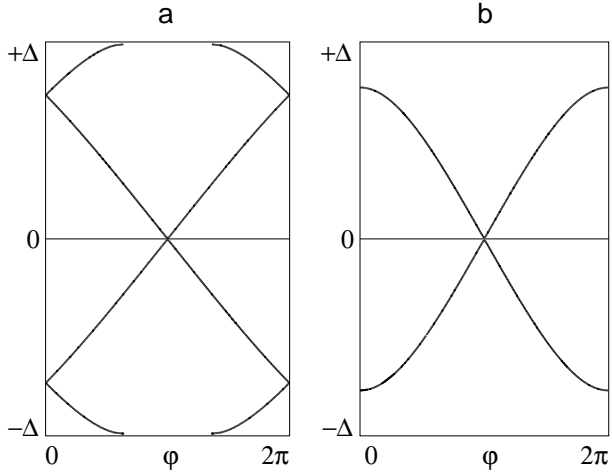}
\caption{ Andreev levels in a single mode $SNS$ junction with $d=\hbar v_F/
\Delta $: (a) $\mathcal{T}_{1,2}=1$ and (b)  $\mathcal{T}_{1,2}\ll 1$
and $\Delta /\Gamma = 0.5$. In both cases $\mathcal{T}=1$.  } 
    \label{fig:alsns-res} 
  \end{center} 
\end{figure} 

In order to understand the physical reasons for such a difference 
it is instructive to compare the structure of discrete Andreev levels for 
ballistic ($\mathcal{T}_{1,2}=1$) $SNS$ junctions with that for $SINIS$
junctions with $\mathcal{T}_{1,2}\ll1$, 
see Fig. 4. The spectrum of an
$SINIS$ system consists only of a single non-degenerate state $\varepsilon _0 
(\varphi)$ in the interval $0 < \varepsilon _0 < \Delta$ (Fig. 4b). As a
result, the behavior of $\varepsilon _0 (\varphi)$ at small $\varphi $ is
smooth and the derivative of
$\varepsilon _0$ with respect to $\varphi $ has no jump at $\varphi =0$.
In contrast, in the case of ballistic $SNS$ junctions discrete levels
become split at arbitrary small values of the phase $\varphi $
(Fig. 4a). Hence, the derivative of the lowest Andreev level with respect to
$\varphi $ acquires a jump at $\varphi =0$. As we have already discussed, this
feature is crucial for the presence of spontaneous currents in superconducting
rings with arbitrary small inductances $\mathcal{L}$ and  
with odd number of electrons. Since this feature is absent in the case
of  $SINIS$ junctions the spontaneous current in the
ground state of the system  can only develop at not very small ring inductances.

Analytical expressions for the energy of the subgap state (obtained from 
the poles of the summand in Eq. (\ref{fbfb1})) and for the parity-dependent
currents $I_{e/o}$ can be derived in various asymptotic regimes. At a given
value of the total transmission $\mathcal{T}$ one can distinguish two 
limiting cases: $\Delta /\Gamma \ll 1$ and  $\Delta /\Gamma \gg 1$.
In the case of a wide resonance $\Delta /\Gamma \ll 1$, the energy of the
subgap state  is $\varepsilon_0(\varphi)=\Delta \sqrt{1-
\mathcal{T}\sin^2(\varphi/2)}$ and $I_{e/o}$ are given by the 
expressions derived
for a single mode QPC with an effective transmission $\mathcal{T}$. In the
opposite narrow resonance limit $\Delta /\Gamma \gg 1$ we get 
$$
\varepsilon_0(\varphi)=\sqrt{(\epsilon_R )^2+\frac{1}{4} \Gamma ^2} \sqrt{1-
\mathcal{T}\sin^2(\varphi/2)}.
$$ 
The currents $I_{e/o}$ are again given by the expressions obtained for 
QPC with the substitution of $\Delta$ by $\sqrt{(\epsilon_R )^2+\frac{1}{4} 
\Gamma ^2}$. Note that in both these asymptotic regimes only the discrete
spectrum contributes to the currents $I_{e/o}$. Thus, for these regimes 
the Josephson current will be completely blocked by the odd electron in the
limit of low temperatures.

At  intermediate values of the parameter $\Delta /\Gamma $ the continuous 
spectrum will contribute to the Josephson current.
The current-phase relations $I_{e/o} (\varphi)$ can be easily evaluated 
numerically for  arbitrary parameter values. The results of these calculations
 -- partially represented in Fig. 5 --  clearly demonstrate that at
 sufficiently low  temperatures the ``$\pi$-junction'' state should be 
realized in the case of odd number of electrons. 
According to our expectations, however, in the case of resonant $SINIS$
junctions the current has no 
jump at $\varphi =0$ and the  behavior of $I_o(\varphi )$ is more
similar to that of ``conventional'' $\pi$-junctions \cite{Leva}.

\begin{figure}[tb] 
\begin{center}
   \includegraphics[angle=0,width=.5\textwidth]{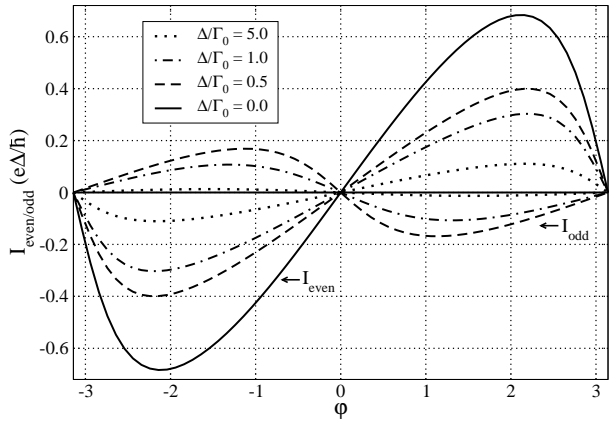}
\caption{ Zero temperature current-phase relations $I_{e/o}(\varphi )$
for $\mathcal{T}=0.9$, $\Gamma_1=\Gamma_2=\Gamma_0$ and different values
of the parameter $\Delta/\Gamma_0$.   } 
    \label{fig:Ieo-res} 
  \end{center} 
\end{figure} 

Let us emphasize again that all the above results for $SINIS$ junctions with
small interface transmissions were derived in the limit of short junctions 
$d \ll \xi_0$. Resonant effects are also important in the opposite limit
$d \gg \xi_0$ \cite{GZ}. However, in the latter case even exactly at 
resonance the Josephson current remains small being proportional to the
smallest of the two transmissions $\mathcal{T}_1$ and $\mathcal{T}_2$. 
Hence, the parity effect in the limit $d \gg \xi_0$ will also
be small in equal measure. For this reason the case of
long junctions with resonant transmission  will not be considered here.

\section{Conclusions}

In this paper we have demonstrated that at sufficiently low temperatures 
superconducting
parity effect may strongly influence equilibrium persistent
currents in isolated superconducting nanorings containing a
weak link with few conducting modes. An odd electron, being added
to the ring, occupies the lowest available Andreev state and
produces a countercurrent circulating inside the ring. For a
single channel quantum point contact at $T=0$ this countercurrent
exactly compensates the supercurrent $I_e$ produced by all other
electrons and, hence, yields complete {\it blocking} of PC for any
value of the external magnetic flux. In superconducting nanorings
with embedded normal metal the odd electron countercurrent can
``overcompensate'' $I_e$ and a novel ``$\pi /N$-junction'' state
occurs in the system. Changing the electron parity number from
even to odd results in {\it spontaneous} supercurrent in the
ground state of such rings without any externally applied magnetic
flux. This and other novel effects predicted here 
can be directly tested in modern experiments.
They can also be used for practical realization of superconducting qubits.

\section*{Acknowledgments}
This work is part of the
Kompetenznetz ``Funktionelle Nanostructuren'' supported by the Landestiftung
Baden-W\"urttemberg gGmbH. This work has also been supported
by the European Community's Framework Programme
NMP4-CT-2003-505457 ULTRA-1D "Experimental and theoretical investigation of
electron transport in ultra-narrow 1-dimensional nanostructures". 


\begin{thebibliography}{10}
\bibitem{AN92} D.V. Averin and Yu.V. Nazarov, Phys.\ Rev.\ Lett. 69 (1992) 1993.
\bibitem{Tuo93} M.T. Tuominen, J.M. Hergenrother, T.S. Tighe, and M. Tinkham,
Phys.\ Rev.\ Lett. 69 (1992) 1997.
\bibitem{Laf93} P. Lafarge, P. Joyez, D. Esteve, C. Urbina, and M.H. Devoret, 
Phys.\ Rev.\ Lett. 70 (1993) 994.
\bibitem{JSA94} B. Janko, A. Smith, and V. Ambegaokar,
Phys.\ Rev.\ B 50 (1994) 1152.
\bibitem{GZ94} D.S. Golubev and A.D. Zaikin,
Phys.\ Lett.\ A195 (1994) 380.
\bibitem{AN94}D.V. Averin and Yu.V. Nazarov, Physica B203 (1994) 310.
\bibitem{SZ} S.V. Sharov and A.D. Zaikin, Phys.\ Rev.\ B 71 (2005) 014518.
\bibitem{Kang} K. Kang, Europhys.\ Lett.\ 51 (2000) 181.
\bibitem{Yak} H.-J. Kwon and V.M. Yakovenko, Phys.\ Rev.\
Lett.\ 89 (2002) 017002.
\bibitem{QPSth} A.D. Zaikin, D.S. Golubev, A. van Otterlo, and G.T. Zim\'anyi,
Phys.\ Rev.\ Lett.\ 78 (1997) 1552.
\bibitem{QPSth2} D.S. Golubev and A.D.
Zaikin, Phys.\ Rev.\ B 64 (2001) 014504.
\bibitem{QPSexp} A. Bezryadin, C.N. Lau, and
M. Tinkham, Nature (London) 404 (2000) 971.
\bibitem{QPSexp2} C.N. Lau, N. Markovic,
M. Bockrath, A. Bezryadin, and M. Tinkham, Phys.\ Rev.\ Lett.\ 87
(2001) 217003.
\bibitem{FN9} The relation (\ref{qpc}) is strictly applicable only for
grand canonical
ensembles. However, at this point the difference between
canonical and grand canonical ensembles is unimportant and can be disregarded.
\bibitem{GZ} A.V. Galaktionov and A.D. Zaikin, Phys.\ Rev.\ B 65 
(2002) 184507.
\bibitem{KO} I.O. Kulik and A.N. Omel'yanchuk, Sov.\ J.\ Low Temp.\ Phys.\
  4 (1978) 142.
\bibitem{HKR} W. Haberkorn, H. Knauer, and J. Richter, Phys.\ Stat.\
Solidi (A) 47 (1978) K161.
\bibitem{Furusaki} A. Furusaki and M. Tsukada, Physica B165-166 (1990) 967.
\bibitem{Bee} C.W.J. Beenakker and H. van Houten, Phys.\ Rev.\ Lett.\ 66
  (1991) 3056.
\bibitem{HE} M. Hayashi and H. Ebisawa, Phys.\ Rev.\ B 67 (2003) 014524.
\bibitem{Kulik} I.O. Kulik, Sov.\ Phys.\ JETP 30 (1970) 944.
\bibitem{Ishii} C. Ishii, Progr.\ Theor.\ Phys.\ 44 (1970) 1525.
\bibitem{Leva} L.N. Bulaevskii, V.V. Kuzii, and A.A. Sobyanin, JETP Lett.\
  25 (1977) 290.
\bibitem{BGZ} Yu.S. Barash, A.V. Galaktionov and A.D. Zaikin, 
Phys.\ Rev.\ B 52 (1995) 665.
\bibitem{Volkov} A.F. Volkov, Phys.\ Rev.\ Lett.\ 74 (1995) 4730.
\bibitem{WSZ} F.K. Wilhelm, G. Sch\"on, and A.D. Zaikin, Phys.\ Rev.\ Lett.\
81 (1998) 1682.
\bibitem{Yip} S.K. Yip, Phys.\ Rev.\  58 (1998) 5803.
\bibitem{FN8} For small $\mathcal{L} \to 0$ non-vanishing spontaneous 
currents in the ground state may as well exist in systems
containing $SNS$ junctions formed by $d$-wave superconductors \cite{BGZ}.  Also in that case this feature is due to the non-sinusoidal
current-phase relation in such junctions.

\end{thebibliography}

\end{document}